\documentclass[twocolumn]{IEEEtran}
\usepackage{amsmath}
\usepackage{amsthm}
\usepackage{amsfonts}
\usepackage{amssymb}
\usepackage{caption}
\usepackage{mathrsfs}
\usepackage{cases}
\usepackage{cite}
\usepackage{graphicx}
\usepackage{mathrsfs}
\usepackage{subfigure}
\usepackage{epstopdf}
\usepackage{color}

\usepackage{enumitem}
\usepackage{setspace}
\usepackage{bm}

\theoremstyle{remark}
\newtheorem{theorem}{\hspace{1em}Theorem}

\begin{document}

\title{\LARGE Cooperative Pilot Spoofing in MU-MIMO Systems
%
\thanks{
The authors are with the School of Information and Communication Engineering, and also with the Ministry of Education Key Lab for Intelligent Networks and Network Security, Xi’an Jiaotong University, Xi’an, 710049, Shaanxi, China (e-mail:xjbswhm@gmail.com; xjtuwsd@stu.xjtu.edu.cn).
}
\author{ Hui-Ming Wang,~\IEEEmembership{Senior Member,~IEEE}, and Shao-Di Wang}}
\maketitle

\pagenumbering{gobble}

\begin{abstract} 
In this letter, we consider downlink transmission of a multiuser multiple-input multiple-output (MU-MIMO) system with zero-forcing (ZF) precoders in the presence of multiple attackers. We propose a cooperative pilot spoofing attack (CPSA), where the attackers collaboratively impair the channel estimations in the uplink channel training phase, aiming at deteriorating the downlink thoughput of the whole cell. We first evaluate the impacts of CPSA on the channel estimation and the downlink ZF precoding design, and then we derive an analytical expression for the achievable downlink sum-rate. Furthermore, we investigate the optimal attack strategy to minimize the achievable downlink sum-rate. We show that the optimization problem under consideration is a convex one so the global optimum could be obtained conveniently. Numerical results show that the CPSA attack results in a severe performance deterioration with the increase in the attacking power and the number of attackers.
\end{abstract}
\begin{IEEEkeywords}
Physical layer security, pilot spoofing, achievable downlink sum-rate, convex optimization.
\end{IEEEkeywords}

\section{Introduction}
\label{Sec:Introduction}
MU-MIMO is the most promising manner of exploiting the spatial degrees of freedom provided by multiple-antennas at the base station (BS) \cite{1}. To fully exploit benefits of MU-MIMO, accurate channel state information (CSI) is a prerequisite. 
In practice, the CSI needs to be estimated. In a time-division duplex (TDD) system, the BS estimates the CSI based on the uplink pilot signals due to the reciprocity of the uplink and downlink channels \cite{3}.

%

However, this specific pilot transmission mechanism is vulnerable to malicious interference from active attackers. In particular, the malicious attacker can attack the uplink pilot transmission by sending the same pilot signals as legitimate users, which is also known as pilot spoofing attack (PSA) \cite{2}, \cite{4}. PSA may lead to incorrect channel estimations and consequently reduce the wireless thoughput of legitimate users in the whole cell significantly.

Recently, PSA has attracted a lot of research interest \cite{5}-\cite{8}.
In \cite{5}, the authors studied the impact of a PSA launched by a single-antenna attacker in a single user scenario, where analysis showed that this attack could drastically weaken the strength of the received signal at the legitimate user. Extreme cases were considered where the number of transmit antennas and the attacker’s power were very large. In \cite{6}, the authors investigated a PSA launched by a multi-antenna attacker in a multi-cell multiuser massive MIMO system, and they found that the attacker could conduct a best possible PSA by maximizing the total average estimation error variance of the desired user’s channel, because the leakage of the desired signal would increase when the channel estimation error increased. In \cite{7}, the authors studied a combined PSA in a single-cell massive MIMO system, and the downlink transmission rates in the presence of the attack was derived by exploiting the channel hardening effect. In \cite{8}, the authors investigated the design of a PSA carried out by multiple single-antenna attackers in a single user scenario. They constructed an optimization problem from the point of view of the attackers, which aimed to maximize the signal-to-noise ratio (SNR) and information leakage to a target adversary.

However, all these aforementioned works are limited either within a single attacker \cite{5,6,7} or focusing the impact on a single specific user \cite{5}, \cite{8}. In fact, on one hand, the PSA may effect all users in the whole cell, which may deteriorate the cell performance severely. On the other hand, since the user access protocol is publicly known, multiple attackers can synchronized to the BS and lauch collaborative attack to improve their PSA capabilities\footnote{In practice, the attackers can be connected to each other via low-cost low-capacity wireless links, so they can share their CSIs for collaboration.}.  
So far, a study on a general PSA scenario with multiple users and multiple attackers is still absent, so the ultimate impact of PSA to a MU-MIMO in a cell for multiple users has not been clearly exposed yet. Although the analysis presented in \cite{7} is in this line but it assumes the channel hardening property so it does not hold for moderately large number of transmit antennas (dozens of antennas), which is a more practical scenario.

In this letter, we consider a  MU-MIMO system under the PSA lauched by multiple cooperative attackers, and investigate how multiple attackers can cooperatively perform the PSA to deteriorate the cell performance. Especially, 1) we first evaluate the impacts of CPSA on the channel estimation and the downlink ZF precoding design, and then we derive an analytical expression for the achievable downlink sum-rate. 2)  Furthermore, we investigate the optimal attack strategy, which aims at minimizing the achievable downlink sum-rate. We show that this problem under consideration is a convex optimization problem so the global optimum could be obtained conveniently. 3) Our results show that the CPSA results in a severe performance deterioration for the
whole cell. Several cooperative attackers could drive the sum-rate down to only 30\% of the normal thoughput without attack.

\section{System Model and Problem Description}
\subsection{System Model}
\begin{figure}[t]
 	\centering
 	\includegraphics[width=2 in]{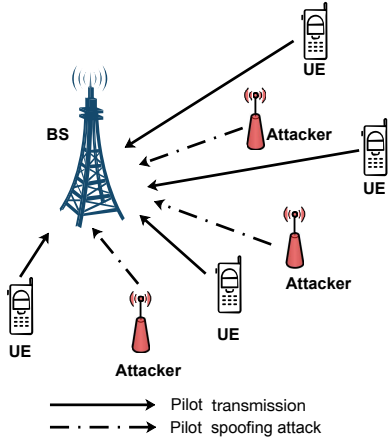}
 	\caption{ System model with multiple cooperative attackers.}
 	\label{w1}
 \end{figure}

We consider a single-cell multiuser TDD communication system, where an $M$-antenna BS serves $K$ single-antenna users using orthogonal pilot sequences for channel training, i.e., $\bm{p}_k \in {\Bbb C}{^{{\tau _p} \times 1}}$ is the pilot sequence of the $k$th user satisfying $\bm{p}_k^T\bm{p}_k^ *  = 1$ with the length $\tau _P$. In this letter, we consider a CPSA, as illustrated in Fig. \ref{w1}, where $N$ single-antenna attackers collaborate to send multiusers pilot sequences combination to disturb the uplink channel training.
Since the initial access and pilot transmission protocol are publicly known \cite{5},  each attacker could easily synchronize with the BS and  replicates the same pilot signals to confound the BS cooperatively. 

We use $\bm{h}_{B,k}^{} \in {{\Bbb C}^{M \times 1}}$ and $\bm{h}_{A,n}^{} \in {{\Bbb C}^{M \times 1}}$ to denote the channel from the $k$th user and the $n$th attacker to BS, respectively. In particular, ${\bm{h}_{B,k}}{\text{ = }}\sqrt {{\beta _{B,k}}} {\bm{g}_{B,k}}$, ${\bm{h}_{A,n}}{\text{ = }}\sqrt {{\beta _{A,n}}} {\bm{g}_{A,n}}$, where ${\beta _{B,k}}$ and ${\beta _{A,n}}$ denote the large-scale fadings, ${\bm{g}_{B,k}} \in {{\Bbb C}^{M \times 1}}$ and ${\bm{g}_{A,n}} \in {{\Bbb C}^{M \times 1}}$ are the small-scale Rayleigh fadings with each element independent and identical distributed (i.i.d.) complex Gaussian random variable with zero mean and unit variance, i.e., ${\bm{g}_{B,k}}\sim\mathbb{C}\mathbb{N}\left( \bm{0},{\bm{I}_{M}} \right)$, ${\bm{g}_{A,n}}\sim\mathbb{C}\mathbb{N}\left( \bm{0},{\bm{I}_{M}} \right)$.
 
\subsection{CPSA Scheme}
In the CPSA scheme, each attacker target all the users so the attack signal is a combination of all users’ pilot signals. Take the case of the $n$th attacker, for $n = 1,2, \cdots ,N$, the attack signal can be expressed as
\begin{align}
{\bm{s}_{{A_n}}} = \sum\limits_{k = 1}^K {\sqrt {{\tau _p}{{\theta} _{n,k}}{P_{{A_n}}}} } {\bm{p}_k},
\end{align}
where ${P_{{A_n}}}$ is the attacking power, ${\theta _{n,k}}$ is the power allocation coefficient allocated for attacking the pilot of the $k$th user for $k = 1,2, \cdots ,K$. We can see  in (1) that the attackers transmit a linear combination of all users’ pilot sequences to confound the BS. Since the pilot information is publicly known, in this way, the CPSA scheme does not need to have any prior knowledge on which pilot sequences are currently used by the legitimate users, which is more practical. 

In the uplink pilot transmission phase, the BS will receive a pilot signal combination from users and attackers as  
\begin{align}
{\bm{Y}_B} = \sum\limits_{k = 1}^K {\sqrt {{\tau _p}{P_{{U_k}}}} \bm{h}_{B,k}^{}} \bm{p}_k^T + \sum\limits_{n = 1}^N {\bm{h}_{A,n}^{}\bm{s}_{{A_n}}^T}  + \bm{U},
\end{align}
where ${P_{{U_k}}}$ is the pilot transmission power of the $k$th user, $\bm{U}$ is the additive Gaussian white noise (AWGN) matrix at the BS with each element being distributed as ${\Bbb C}{\Bbb N}\left( {{0},{{\sigma ^{\text{2}}}}} \right)$.

As these pilot sequences are orthogonal to each other, i.e., $\bm{p}_i^H\bm{p}_k=0$ for $i\neq k$, the BS can obtain the channel estimation result of the $k$th user under a priori knowledge of $\bm{p}_k^ * $ by the following pre-processing
\begin{align}
 {\bm{y}_{B,k}}&={{\left( {{\bm{Y}_B}\bm{p}_k^ * } \right)} \mathord{\left/
 {\vphantom {{\left( {{Y_B}*\bm{p}_k^ * } \right)} {\left( {\sqrt {{\tau _p}{P_{{U_k}}}} } \right)}}} \right.
 \kern-\nulldelimiterspace} {\left( {\sqrt {{\tau _p}{P_{{U_k}}}} } \right)}} \nonumber\\
  & ={\bm{h} _{B,k}} + \sum\limits_{n = 1}^N \sqrt {{\theta _{n,k}} \frac{P_{A_n}}{P_{U_k}}}  {\bm{h}_{A,n}} + \frac{ \bm{z}_{B,k}} { {\sqrt {{\tau _p}{P_{{U_k}}}} }},
\end{align}
where ${ \bm{z}_{B,k}}$ is the equivalent noise vector with distribution ${\Bbb C}{\Bbb N}\left( {\bm{0},{{{\sigma ^{\text{2}}}}\bm{I}_M}} \right)$. Without awaring the CPSA attack, the BS will calculate the MMSE estimation of the legitimate channel ${\bm{h}_{B,k}}$ via the standared process as \cite{9}
\begin{align}
 {\bm{\hat h}_{B,k}}  & \mathop {=}{\bm{\Psi}_{B,k}}\bm{\Omega} _{B,k}^{ - 1}{\bm{y}_{B,k}}, 
\end{align}
where ${\bm{\Psi}_{B,k}}\triangleq {\Bbb E}\left\{ {\bm{h}_{B,k}^{}\bm{y}_{B,k}^H} \right\}$, and $\bm{\Omega} _{B,k}^{ - 1}\triangleq{\Bbb E}{\left\{ {\bm{y}_{B,k}^{}\bm{y}_{B,k}^H} \right\}^{ - 1}}$
are the covariance matrices, which can be derived as
\begin{align}
\nonumber
  & {\bm{\Psi} _{B,k}} = {\beta _{B,k}}{\bm{I}_M},  \cr 
  & \bm{\Omega}_{B,k}^{}  = {\beta _{B,k}}{\bm{I}_M} + \sum\limits_{n = 1}^N {{\theta _{n,k}}\left( {{{{P_{{A_n}}}} \mathord{\left/
 {\vphantom {{{P_{{A_n}}}} {{P_{{U_k}}}}}} \right.
 \kern-\nulldelimiterspace} {{P_{{U_k}}}}}} \right)} {\beta _{A,n}}{\bm{I}_M} \cr 
  & {\text{         }} + \left( {{{{\sigma ^2}} \mathord{\left/
 {\vphantom {{{\sigma ^2}} {{\tau _p}{P_{{U_k}}}}}} \right.
 \kern-\nulldelimiterspace} {{\tau _p}{P_{{U_k}}}}}} \right){\bm{I}_M}, 
\end{align}

The estimated channel vector ${\bm{\hat h}_{B,k}}$ is distributed as $\mathbb{C}\mathbb{N}\left( \bm{0},{{\bm{\hat R}_{B,k}}}\right)$ with ${{\bm{\hat R}_{B,k}}}$ can be written as
\begin{align}
{\bm{\hat R}_{B,k}} = \bm{\Psi }_{B,k}^H\bm{\Omega}_{B,k}^{ - 1}\bm{\Psi }_{B,k}^{} = {{\lambda }_{B,k}}{\bm{I}_{M}},
\end{align}
where 
\begin{align}
\nonumber
{\lambda _{B,k}}=\frac{{{\tau _p}{P_{{U_k}}}\beta _{B,k}^2}}{{{\tau _p}{P_{{U_k}}}{\beta _{B,k}}{\text{ + }}{\tau _p}\sum\limits_{n = 1}^N {{\theta _{n,k}}{P_{{A_n}}}} {\beta _{A,n}}{\text{ + }}{\sigma ^{\text{2}}}}}.
\end{align}

The uncorrelated channel estimation error ${\bm{\tilde{h}}_{B,k}}$  satisfing $\bm{{h}}_{B,k}=\bm{\hat{h}}_{B,k}+\bm{\tilde{h}}_{B,k}$  can be derived by invoking the orthogonality property of MMSE estimation as
\begin{align}
{\bm{\tilde{h}}_{B,k}}\sim\mathbb{C}\mathbb{N}\left( \bm{0},{\eta _{B,k}}{\bm{I}_{M}} \right),
\end{align}
where ${{\eta }_{B,k}}\triangleq {{\beta }_{B,k}}-{{\lambda }_{B,k}}$.

\emph{Remark 1:} Note that the attackers can also transmit Gaussian random interference to degrade the accuracy of the channel estimation. However, as shown in our previous works \cite{8}, \cite{10},  transmitting random interference can not offer any advantage over the proposed pilot spoofing signals, which will be shown later in numerical results.

\section{Downlink Sum-Rate Analysis And Optimal Attack Strategy}
\subsection{Downlink ZF Beamforming}
Since ZF downlink beamforming is an asympototically optimal solution for MU-MISO transmission\footnote{ZF precoding can achieve asymptotically optimal throughput in the downlink of MU-MISO system, which has been proved in \cite{11}, \cite{12}.}
Here we consider ZF beamformer for the BS, which is
\begin{align}
\bm{w}_k\mathop  = \limits^\Delta  \frac{{{\bm{a}_{B,k}}}}{{\left\| {{\bm{a}_{B,k}}} \right\|}},
\end{align}
for the $k$th user, where ${\bm{a}_{B,k}}$ is the $k$th column of ${\bm{\hat H}_B}{\left( {{{\bm{\hat H}}^H}_B{\bm{\hat H}_B}} \right)^{ - 1}}$, and ${\bm{\hat H}_B} \triangleq[\bm{\hat h}_{B,1}^{}, \cdots ,\bm{\hat h}_{B,K}^{}]$ is the channel estimation matrix. Due to the CPSA, the BS uses the impaired ZF precoder for downlink data transmission. The received signal at the $k$th user can be written as 
\begin{align}
{y_k} = \sqrt {{P_{{B_k}}}} \bm{h} _{B,k}^H\bm{w}_k^{}{s_k} + \sum\limits_{i = 1,i \ne k}^K {\sqrt {{P_{{B_i}}}} \bm{h} _{B,k}^H\bm{w}_i^{}{s_i} + {{z}_k}}, 
\end{align}
where ${P_{{B_k}}}$ is the transmit power allocated for the $k$th user, and ${{z}_{k}}\sim\mathbb{C}\mathbb{N}\left( {0},{\sigma ^{\text{2}}} \right)$ is the additive noise at the $k$th user.

As discussed in \cite{13}, without the dedicate downlink channel training, the users only have statistical effective channel gain for signal demodulation, and the signal received at the $k$th user can be reformulated as\footnote{In the absence of downlink channel training, statistical CSI is used by each user for signal detection. This is a standard detection scheme, and has been widely adopted in \cite{13,14,15}.}
\begin{align}
& {y_k} = \sqrt {{P_{{B_k}}}} {\Bbb E}\left\{ {\bm{h} _{B,k}^H\bm{w}_k^{}} \right\}s_k^{}  \cr 
  & \qquad + \sqrt {{P_{{B_k}}}} \left( {\bm{h} _{B,k}^H\bm{w}_k^{} - {\Bbb E}\left\{ {\bm{h} _{B,k}^H\bm{w}_k^{}} \right\}} \right)s_k^{}  \cr 
  & \qquad  + \sum\limits_{i = 1,i \ne k}^K {\sqrt {{P_{{B_i}}}} \bm{h} _{B,k}^H\bm{w}_i^{}s_i^{} + z_k^{}},
\end{align}
where due to the incorrect ZF precoding caused by CPSA, inter-user interference occurs, which will greatly deteriorate the overall throughput in the cell.

\subsection{Achievable Downlink Sum-Rate}
The achievable downlink rate at the $k$ user can be given by
\begin{align} 
{R_k} = \log\left( {1 + {\gamma _k}} \right), \label{R}
\end{align}
where
\begin{align}
\nonumber
{\gamma _k} = \frac{{{P_{{B_k}}}{{\left| {{\Bbb E}\left\{ {\bm{ h}_{B,k}^H\bm{w}_k^{}} \right\}} \right|}^2}}}{{{P_{{B_k}}}{\Bbb V}{\text{ar}}\left\{ {\bm{h}_{B,k}^H\bm{w}_k^{}} \right\} + \sum\limits_{i = 1,i \ne k}^K {{P_{{B_i}}}{\Bbb E}\left\{ {{{\left| {\bm{h}_{B,k}^H\bm{w}_i^{}} \right|}^2}} \right\}}  + {\sigma ^{\text{2}}}}},
\end{align}
and ${\Bbb E}\left\{  \cdot  \right\}$ and ${\Bbb V}{\text{ar}}\left\{  \cdot  \right\}$ are the expectation and variance operators, respectively.

To simplify the subsequent analysis, we assume in the uplink pilot transmission phase ${P_U}=P_{U_k}$, and ${P_A}={P_{{A_n}}}$ for all $k$ and $n$. In addition, in the downlink data transmission phase, ${P_B}={P_{{B_k}}}$. Then, the achievable downlink rate at the $k$th user can be derived as follows.

\begin{theorem}
Under the CPSA and  ZF precoding, the achievable downlink rate at the $k$th user is 
\begin{align}
 {\tilde R_k}  =\log\left( {1 + \frac{{{A_k}}}{{{B_k} + {C_k}{\bm{\nu} ^T}{\bm{\theta}  _k}}}} \right),
\end{align}
where
${\bm{\theta} _k} \triangleq {\left[ {{\theta _{1,k}}, \cdots ,{\theta _{N,k}}} \right]^T}$, $\bm{\nu} \triangleq [{\beta _{A,1}}, \cdots ,{\beta _{A,N}}{]^T}$, ${A_k} \triangleq \xi \left( {M - K{\text{ + }}1} \right){\tau _p}{P_U}\beta _{B,k}^{\text{2}}$, ${B_k} \triangleq \left( {M - 2K + 1 - \xi \left( {M- } \right.} \right.$\\
$\left. {\left. { - K{\text{ + }}1} \right)} \right){\tau _p}{P_U}\beta _{B,k}^{\text{2}} + \left( {K{\beta _{B,k}} + {\sigma ^{\text{2}} \mathord{\left/{\vphantom {1 {{P_B}}}} \right.\kern-\nulldelimiterspace} {{P_B}}}} \right)\left( {{\tau _p}{P_U}{\beta _{B,k}} + \sigma ^{\text{2}}} \right)$,\\
${C_k} \triangleq \left( {K{\beta _{B,k}} + {\sigma ^{\text{2}} \mathord{\left/{\vphantom {1 {{P_B}}}} \right.\kern-\nulldelimiterspace} {{P_B}}}} \right){\tau _p}{P_A}$, $\xi \left( x \right) \triangleq {{\Gamma \left( {x + 1/2} \right)} \mathord{\left/ {\vphantom {{\Gamma \left( {x + 1/2} \right)} {\Gamma \left( x \right)}}} \right.\kern-\nulldelimiterspace} {\Gamma \left( x \right)}}$.
\end{theorem}
\emph{Proof:} By calculating the following three terms in ${\gamma _k}$ in (\ref{R}), the derivation of the achievable downlink rate at the $k$th user is outlined.

For the numerator  ${\left | {{\Bbb E}\left\{ {\bm{h} _{B,k}^H\bm{w}_k^{}} \right\}} \right|^2}$, it can be calculated by
\begin{align}
{\left| {{\Bbb E}\left\{ {\bm{h}_{B,k}^H\bm{w}_k^{}} \right\}} \right|^2}&\mathop {\text{ = }}\limits^{(a)} {\left| {{\Bbb E}\left\{ {\left( {\bm{\hat h}_{B,k}^H{\text{ + }}\bm{\tilde{h}}_{B,k}^H} \right)\bm{w}_k^{}} \right\}} \right|^2} 
\nonumber \\
&\mathop {\text{ = }}\limits^{(b)} {\left| {{\Bbb E}\left\{ {{1 \mathord{\left/
 {\vphantom {1 {\left\| {{\bm{a}_{B,k}}} \right\|}}} \right.
 \kern-\nulldelimiterspace} {\left\| {{\bm{a}_{B,k}}} \right\|}}} \right\}} \right|^2}
 \nonumber \\
 &\mathop {\text{ = }}\limits^{(c)} \xi \left( {M - K{\text{ + }}1} \right){\lambda _{B,k}},
\end{align}
where $\xi \left( x \right) \triangleq {{\Gamma \left( {x + 1/2} \right)} \mathord{\left/
 {\vphantom {{\Gamma \left( {x + 1/2} \right)} {\Gamma \left( x \right)}}} \right.
 \kern-\nulldelimiterspace} {\Gamma \left( x \right)}}$, step ($a$) is obtained by applying the MMSE channel estimation error model, step ($b$) holds since $\bm{\tilde{h}}_{B,k}^{}$ and $\bm{w}_k^{}$ are uncorrelated, and $\bm{\hat h}_{B,k}^H\bm{w}_k^{} = {1 \mathord{\left/
 {\vphantom {1 {\left\| {{\bm{a}_{B,k}}} \right\|}}} \right.
 \kern-\nulldelimiterspace} {\left\| {{\bm{a}_{B,k}}} \right\|}}$, and step ($c$) results from the  Gamma distribution.

${\Bbb V}{\text{ar}}\left\{ {\bm{h}_{B,k}^H\bm{w}_k^{}} \right\}$ in denominator can be computed by
\begin{align} 
  & {\Bbb V}{\text{ar}}\left\{ {\bm{h}_{B,k}^H\bm{w}_k^{}} \right\}\mathop {\text{ = }}\limits^{(a)} \mathbb{E}\left\{ {{{\left| {\bm{h}_{B,k}^H\bm{w}_k^{}} \right|}^2}} \right\} - {\left| {\mathbb{E}\left\{ {\bm{h}_{B,k}^H\bm{w}_k^{}} \right\}} \right|^2}  
  \nonumber \\ 
  & {\text{}}\mathop {\text{ = }}\limits^{(b)}  \mathbb{E}\left\{ {{{\left| {\bm{\hat h}_{B,k}^H\bm{w}_k^{}} \right|}^2}} \right\}{\text{ + }}\mathbb{E}\left\{ {{{\left| {\bm{\tilde h}_{B,k}^H\bm{w}_k^{}} \right|}^2}} \right\} - {\left| {\mathbb{E}\left\{ {\bm{\hat h}_{B,k}^H\bm{w}_k^{}} \right\}} \right|^2}
  \nonumber \\
  & {\text{                      }}\mathop  = \limits^{(c)} \mathbb{E}\left\{ {{{\left| {\bm{\tilde h}_{B,k}^H\bm{w}_k^{}} \right|}^2}} \right\}{\text{ + }}\mathbb{V}{\text{ar}}\left\{ {\bm{\hat h}_{B,k}^H\bm{w}_k^{}} \right\}   
  \nonumber \\
  & {\text{                      }}\mathop  = \limits^{(d)} {\eta _{B,k}}{\text{ + }}\left( {M - K + 1 - \xi \left( {M - K{\text{ + }}1} \right)} \right){\lambda _{B,k}}, 
\end{align}
where step ($a$) is obtained by applying the definition of variance, step ($b$) holds since $\bm{\hat h}_{B,k}$ and $\bm{\tilde h}_{B,k}$ are independent of each other, step ($c$) is obtained by applying the definition of variance, and step ($d$) holds since $\bm{\tilde h}_{B,k}$ and $\bm{w}_k^{}$ are uncorrelated.

The term $\sum\limits_{i = 1,i \ne k}^K {{\Bbb E}\left\{ {{{\left| {\bm{h}_{B,k}^H\bm{w}_i^{}} \right|}^2}} \right\}} $ in denominator is
\begin{align}
& \sum\limits_{i = 1,i \ne k}^K {\mathbb{E}\left\{ {{{\left| {\bm{h}_{B,k}^H\bm{w}_i^{}} \right|}^2}} \right\}} \mathop {\text{ = }}\limits^{} \sum\limits_{i = 1,i \ne k}^K {\mathbb{E}\left\{ {{{\left| {\left( {\bm{\hat h}_{B,k}^H{\text{ + }}\bm{\tilde h}_{B,k}^H} \right)\bm{w}_i^{}} \right|}^2}} \right\}}   
\nonumber \\
& {\text{                             }}\mathop {\text{ = }}\limits^{(a)} \sum\limits_{i = 1,i \ne k}^K {\left[ {\mathbb{E}\left\{ {{{\left| {\bm{\hat h}_{B,k}^H\bm{w}_i^{}} \right|}^2}} \right\} + \mathbb{E}\left\{ {{{\left| {\bm{\tilde h}_{B,k}^H\bm{w}_i^{}} \right|}^2}} \right\}} \right]}
\nonumber \\
& \mathop {\text{ = }}\limits^{({\text{b}})} \sum\limits_{i = 1,i \ne k}^K {\mathbb{E}\left\{ {{{\left| {\bm{\tilde h}_{B,k}^H\bm{w}_i^{}} \right|}^2}} \right\}} \mathop {\text{ = }}\limits^{} \left( {K - 1} \right){\eta _{B,k}},  
 \end{align}
where step ($a$) results form the independence of $\bm{\hat h}_{B,k}$ and $\bm{\tilde h}_{B,k}$, and step ($b$) holds since $\bm{\hat h}_{B,k}^H\bm{w}_i^{} = 0$ for $i \ne k$.

Substituing (5), (12), (13) and (14) into (10) yields the expression (11). This completes the proof. $\hfill\square$ 

Accordingly, the achievable downlink sum-rate is
\begin{align}
\nonumber
{R_{sum}} = \sum\limits_{k = 1}^K {{{\tilde R}_k}}  = \sum\limits_{k = 1}^K {\log\left( {1 + \frac{{{A_k}}}{{{B_k} + {C_k}{\bm{\nu }^T}{\bm{\theta }_k}}}} \right).} 
\end{align}
\emph{Remark 2:} Note that the achievable downlink sum-rate does not depend on the small-scale fading components ${\bm{g}_{B,k}}$ and ${\bm{g}_{A,n}}$, which implies that the attackers could optimize the attack without the legitimate CSI. This makes CPSA a more practical attacking scheme.

\subsection{Optimal Attack Strategy}
The goal of the CPSA is to minimize the achievable downlink sum-rate of the target cell by allocating the attacking power. This strategy could be formulated as follows
\begin{align}
\begin{gathered}
   \min\limits_{{\bm{\theta}_k}} {\text{  }}\sum\limits_{k = 1}^K {\log\left( {1 + \frac{{{A_k}}}{{{B_k} + {C_k}{{\bm{\nu}} ^T}{{\bm{\theta}_k}}}}} \right)} , \hfill \\
  s.t.{\text{   C1: }}0 \leqslant {\theta _{n,k}} \leqslant 1,{\text{ }}n = 1,2, \cdots ,N, \hfill \\
  \quad\;{\text{       C2:  }}\sum\limits_{k = 1}^K {{\theta _{n,k}} \leqslant 1}  \hfill \\ 
\end{gathered} 
\end{align}
where the constraints C1 and C2 account for the attacking power sum and individual constraints.

Fortunately, we declear that the optimization problem is a convex problem. 
Denote $f\left( {{\bm{\theta}_k}} \right)\triangleq{{{B_k}} \mathord{\left/{\vphantom {{{B_k}} {{A_k}}}} \right.\kern-\nulldelimiterspace} {{A_k}}} + \left( {{{{C_k}} \mathord{\left/
 {\vphantom {{{C_k}} {{A_k}}}} \right.\kern-\nulldelimiterspace} {{A_k}}}} \right){\bm{\nu} ^T}{\bm{\theta}  _k}$. The first and second derivative of $f\left( {{\bm{\theta}  _k}} \right)$ can be derived as $\frac{{df\left( {{\bm{\theta}_k}} \right)}}{{d{\bm{\theta}_k}}} = \left( {{{{C_k}} \mathord{\left/{\vphantom {{{C_k}} {{A_k}}}} \right.\kern-\nulldelimiterspace} {{A_k}}}} \right)\bm\nu $ and $\frac{{{d^2}f\left( {{\bm{\theta}_k}} \right)}}{{d{\bm{\theta}_k}d{\bm{\theta}}_k^T}} = {\bm0_N}$, respectively, where ${\bm0_N}$ denotes a $N \times N$ null matrix. According to the necessary and sufficient condition of convex function identification, $f\left( {{\bm{\theta}  _k}} \right)$ is a convex function. Due to $\log\left( {1 + {{\text{1}} \mathord{\left/{\vphantom {{\text{1}} x}} \right.\kern-\nulldelimiterspace} x}} \right)$ is convex, we conclude that the composite function $\log\left( {1 + {{\text{1}} \mathord{\left/{\vphantom {{\text{1}} {f\left( {{\bm{\theta}_k}} \right)}}} \right.\kern-\nulldelimiterspace} {f\left( {{\bm{\theta}_k}} \right)}}} \right)$ is a convex function of ${\bm{\theta}_k}$. Considering that the summation of convex functions is convex, C1 and C2 are convex sets, we can proof the optimization problem is convex. So, it can be efficiently solved by standard convex optimization techniques.
 
\section{Numerical Results}
We evaluate the impact of CPSA to the achievable downlink sum-rate through numerical results. We use $\beta={L_0}{d^{ - \alpha }}$ to model the path loss and shadowing fading, where $d$ is the distance between the BS and the user, ${L_0} =-45$ dB and $\alpha  = 3.7$ is the path loss exponent. The users and attackers are uniformly distributed in a circular cell. The inner radius is 50m, the maximum distance of the users is 400m and that of the attackers is $D_A^{\max }$. We consider communication over a 20 MHz bandwidth with noise floor of -90 dBm. We set ${P_U} = 10$ dBm, ${P_A} = 10$ dBm and ${P_B} = 40$ dBm. We set the pilot sequence length ${\tau _P} = K$ symbol durations. The results are averaged over 10000 Monte-Carlo (MC) tests.

\begin{figure}[t]
  \centering
  \includegraphics[width=3 in]{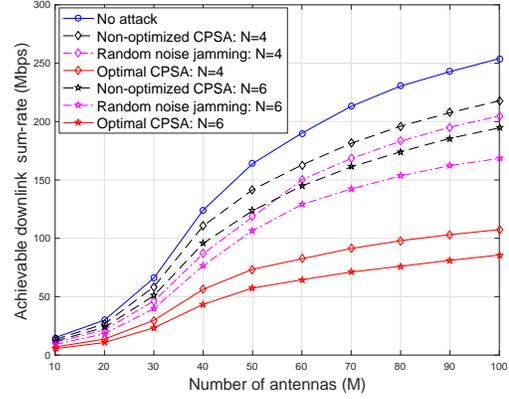}\\
  \caption{The downlink achievable sum-rate versus the number of antennas, where $K=24, D_A^{\max }=300$.}\label{w2}
\end{figure}
Fig. \ref{w2} depicts the achievable downlink sum-rate versus the number of the BS antennas with different number of the attackers. It shows that the achievable downlink sum-rate is significantly degraded by the optimal CPSA. As the number of the attackers increases, the impact becomes more significant. Moreover, we evaluate the impacts of random noise jamming attack (i.e., Gaussian random vector with distribution ${\Bbb C}{\Bbb N}\left( {\bm{0},{\bm{I}_{{\tau _P}}}} \right)$) and non-optimized CSPA (i.e., the attacking power allocated to each user is the same without optimization) as benchmarks. Compared with random noise jamming attack and the non-optimized CSPA, the achievable downlink sum-rate is significantly reduced under the optimal CPSA. These illustrate that the CSPA has severe impact on the CSI estimation and consequently results in a substantial thoughput loss.

\begin{figure}[t]
  \centering
  \includegraphics[width=3 in]{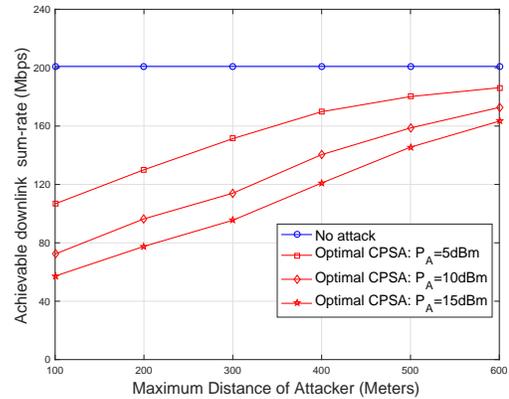}\\
  \caption{The achievable downlink sum-rate versus the maximum distance of the attackers, where $M=64, K=24, N=2$.}\label{w3}
\end{figure}

The impacts of maximum distance $D_A^{\max }$ and the attacking power $P_A$ of the attackers  on the cell thoughput are illustrated in Fig. \ref{w3}. The achievable downlink sum-rate when there is no attack is taken as a benchmark. We observe that when the attacks are not so far away from the BS, the achievable downlink sum-rate has a dramatical deterioration, even there is only two attackers each with power 5 dBm. In addition, increasing attack power a little bit also has severely impact on the whole cell performance.

\section{Conclusion}
\label{Sec:Conclusion}
In this letter, we analyzed the impact of CPSA, i.e., a PSA launched by multiple cooperative attackers. This attack caused a great impact on the channel estimation in the uplink channel training phase. We have evaluated the effect of the CPSA on the achievable downlink sum-rate in a single-cell MU-MIMO system. We shown that the cooperation among attackers can significantly improve their offensive capabilities, and impose dramatic harm to the system throughput. Moreover, it should be noted that most existing pilot spoofing attack detection methods are difficult to be used directly for the CPSA in a MU-MIMO system. For example, random modulation based methods (e.g., random frequency shift [16]) will incur high computational complexity for the MU-MIMO system; it is challenging for statistic feature based methods (e.g., sparsity of virtual channel [17]) to select the optimal detection threshold in the face of such cooperative attacks, and they need to estimate more complicated statistic features when facing the scenario of multiuser and multiple attackers. Consequently, effective detection and defense mechanisms are urgently needed, which is a critical issue for our future research.

\end{document}